\documentclass[letterpaper, 10 pt, conference]{ieeeconf}  

\IEEEoverridecommandlockouts                              

\usepackage{mathrsfs}
\usepackage{graphicx}
\usepackage[T1]{fontenc}
\usepackage{amsmath}
\usepackage{breqn}
\usepackage[utf8]{inputenc}
\usepackage{todonotes}
\usepackage{amssymb}
\usepackage{bm}
\usepackage{url}
\usepackage{cite}

\usepackage{hyperref}
\hypersetup{
    colorlinks=true,
    linkcolor=black,
    filecolor=black,      
    urlcolor=blue,
    pdftitle={Overleaf Example},
    pdfpagemode=FullScreen,
    }

\def\mline{\vrule width4pt height2.5pt depth -2pt}
\def\dashed{\mline\hskip3.5pt\mline}
\def\bdot{\raise.2em\hbox to .15em{.}}

\def\todo#1{\textcolor{black}{#1}}

\title{\LARGE \bf
Upper bound of transient growth in accelerating and decelerating wall-driven flows using the Lyapunov method
}

\author{Zhengyang Wei$^{1}$, Weichen Zhao$^{2}$, and Chang Liu$^{1}$
\thanks{$^{1}$School of Mechanical, Aerospace, and Manufacturing Engineering,
 University of Connecticut, Storrs, CT 06269, USA
        {\tt\small zhengyang.wei@uconn.edu} and {\tt\small chang\_liu@uconn.edu} }%
\thanks{$^{2}$Department of Mechanical Engineering, Binghamton University,
        Binghamton, NY 13902, USA
        {\tt\small wzhao9@binghamton.edu}}%
\thanks{This research was supported by the University of Connecticut (UConn) Research Excellence Program (REP). Z.W. acknowledges the support from the UConn Summer Undergraduate Research Fund (SURF) Awards and the University Scholar Program. The computational work for this project was conducted using resources provided by the Storrs High-Performance Computing (HPC) cluster. We extend our gratitude to the UConn Storrs HPC and its team for their resources and support, which aided in achieving these results. The data and code of this paper are publicly available at \href{https://doi.org/10.5281/zenodo.16876759}{https://doi.org/10.5281/zenodo.16876759}.}
}
\begin{document}

\maketitle
\thispagestyle{empty}
\pagestyle{empty}

\begin{abstract}

This work analyzes accelerating and decelerating wall-driven flows by quantifying the upper bound of transient energy growth using a Lyapunov-type approach. By formulating the linearized Navier-Stokes equations as a linear time-varying system and constructing a time-dependent Lyapunov function, we obtain an upper bound on transient energy growth by solving linear matrix inequalities. This Lyapunov method can obtain the upper bound of transient energy growth that closely matches transient growth computed via the singular value decomposition of the state-transition matrix of linear time-varying systems. Our analysis captures that decelerating base flows exhibit significantly larger transient growth compared with accelerating flows. Our Lyapunov method offers the advantages of providing a certificate of uniform stability and an invariant set to bound the solution trajectory. 

\end{abstract}

\section{Introduction}

Accelerating and decelerating wall-bounded shear flows are frequently encountered in aerospace (e.g., aircraft takeoff and landing), automotive, and industrial processes, where controlling flow stability and transition to turbulence is of great importance \cite{Linot_Schmid_Taira_2024,linot2025extractingdominantdynamicsunsteady,brunton2015closed}. Although linear stability theory has been extensively used to predict the onset of instability in steady flows \cite{schimd2001stability} and periodic flows \cite{floquet1883equations,davis1976stability}, it often underestimates transient energy growth that can lead to substantial amplification of disturbance and potential transition to turbulence \cite{trefethen1993hydrodynamic,butler1992three,schimd2001stability}. This effect is especially pronounced in time-varying shear flows with accelerating and decelerating base flows \cite{Linot_Schmid_Taira_2024,linot2025extractingdominantdynamicsunsteady}, where traditional linear stability analysis fails to fully capture the dynamics. 

The transient growth analysis goes beyond classical linear stability analysis and is critical for understanding the evolution and amplification of initial disturbances \cite{schimd2001stability}. Linear optimal perturbations computed by singular value decomposition (SVD) of the linearized Navier–Stokes operators show that the transient growth of streamwise‐invariant vortices in steady Hagen–Poiseuille flow is proportional to $\text{Re}^2$  \cite{schmid1994optimal}, where Re is the Reynolds number. This reveals a potent transient growth mechanism absent from linear stability analysis. Recently, transient growth in decelerating flow is shown to reach $\mathcal{O}(10^5)$ and scales as $10^{\text{Re}}$ at high Reynolds numbers \cite{Linot_Schmid_Taira_2024}, which will grow much faster over Reynolds number than the scaling law $\text{Re}^2$ for steady base flows. This transient growth in decelerating flow is driven by the Orr mechanism \cite{orr1907stability} unique to the deceleration period, whereas accelerating flows never exceed the transient growth associated with steady base flows \cite{Linot_Schmid_Taira_2024}. Moreover, pulsatile and oscillatory pipe flows show that transient growth of disturbances can be triggered during flow deceleration, which is closely tied to the Stokes‐layer dynamics \cite{xu2021non}. However, existing transient growth analysis of time-varying flows is not tied to the notion of uniform stability and does not provide an invariant set to bound the solution trajectory.

The Lyapunov-based approach can be applied to certify an upper bound on the transient energy growth \cite{kalur2021nonlinear,whidborne2007minimization}, uniform stability for time-varying systems \cite{khalil2002nonlinear,Boyd1994}, and analyze instabilities of a linear time-periodic system \cite{kochnev2025stability}. Such an approach will search for either a common Lyapunov function or a time-varying Lyapunov function by solving Linear Matrix Inequalities (LMI) \cite{Chesi2004, Monta2003,Liu2020, kalur2022estimating, wei2025nonlinear}. For example, homogeneous polynomial Lyapunov functions \cite{Chesi2004} and affine parameter-dependent formulations \cite{Monta2003} enable the expression of stability conditions as LMI for polynomially time-varying systems \cite{Jetto2009}. Moreover, using a time-varying Lyapunov function $V(\boldsymbol{x},t)=\boldsymbol{x}^*\boldsymbol{P}(t)\boldsymbol{x}$ with $\boldsymbol{x}$ as the state variable and $(\cdot)^*$ as a Hermitian operator, we can explicitly incorporate the time-derivative of the Lyapunov matrix $\dot{\boldsymbol{P}}$ in computing $\dot{V}$ to account for the time-varying nature of the system enabling less conservative transient growth and stability certificates.

In this work, we apply this Lyapunov-based approach to obtain an upper bound of transient growth and certify uniform stability in accelerating and decelerating wall-driven flows. We consider a linear time-varying system based on Navier-Stokes equations linearized around a time-varying laminar base flow and then formulate the LMI using time-dependent Lyapunov functions in the form of $V(\boldsymbol{x},t)=\boldsymbol{x}^*\boldsymbol{P}(t)\boldsymbol{x}$ to certify the upper bound of transient energy growth. The time-derivative term $\dot{\boldsymbol{P}}$ is approximated by the forward Euler method. Our Lyapunov-based upper bound of transient growth is then validated to be consistent with the transient growth obtained by SVD of the state-transition matrix \cite{Linot_Schmid_Taira_2024}. Our Lyapunov method systematically captures the enhanced transient growth during deceleration phases identified in recent studies \cite{Linot_Schmid_Taira_2024}. The analysis of eigenvectors of the Lyapunov matrix at initial time $\boldsymbol{P}(t_0)$ reveals that the primary mode aligns in the opposite direction of the laminar base flow profiles, which is consistent with the role of the Orr mechanism \cite{orr1907stability} in amplifying disturbances. Our Lyapunov function $V(\boldsymbol{x},t)$ also certifies uniform stability and provides an invariant set to bound the solution trajectory in more detail compared with transient growth analysis.

This paper is organized as follows: In \S \ref{sec:Problem}, we present the time-varying laminar base flow for accelerating and decelerating wall-driven flows. Section \ref{sec:stability} presents the linearized Navier-Stokes equations and computation of transient growth based on SVD of the state-transition matrix. Section \ref{sec:upper_bounds} then formulates our LMI based on the Lyapunov method to obtain an upper bound of transient growth. We present our results in \S \ref{sec:results} and then conclude this paper in \S \ref{sec:conclusions}.

\section{\label{sec:Problem}Problem Setup and Laminar Base Flows}

\begin{figure}[!htbp]
    \centering
    \includegraphics[width=0.46\textwidth]{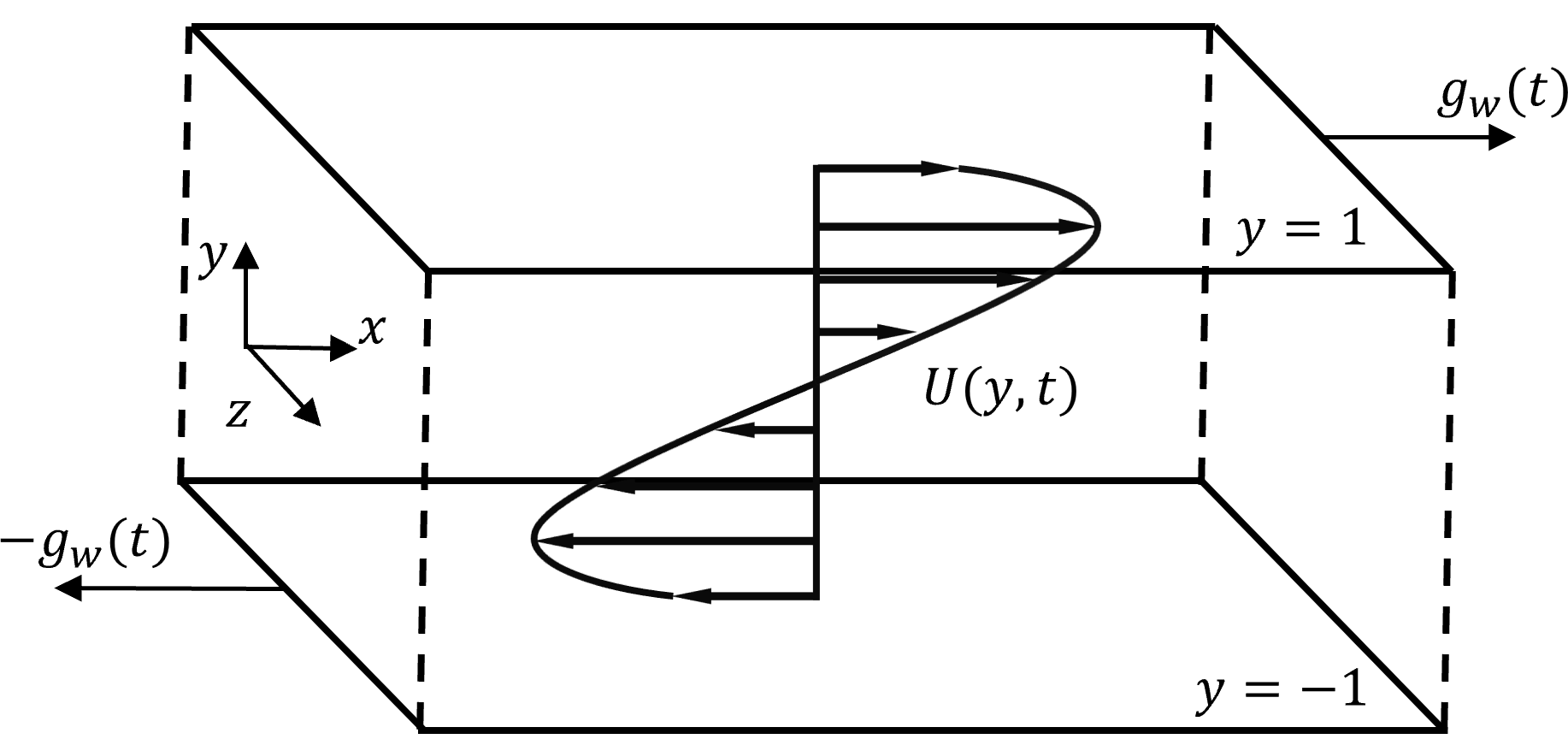}
    \caption{Diagram of wall-driven channel flow, with an example snapshot of the laminar flow for exponentially decaying wall motion.}
    \label{fig1}
\end{figure}

We consider incompressible flow between two parallel moving walls (Fig. \ref{fig1}) with spatial coordinates as streamwise $x \in[-\infty, \infty]$, wall-normal $y \in[-1,1]$, and spanwise $z \in[-\infty, \infty]$ directions. The flow satisfies the incompressible Navier-Stokes equations:
\begin{align}
\partial_t \boldsymbol{u}+\boldsymbol{u} {\cdot} \nabla \boldsymbol{u}=-\nabla p+ \nabla^2 \boldsymbol{u}/\text{Re}, \quad \nabla {\cdot} \boldsymbol{u}=0,
\label{eq:NS}
\end{align}
where $\boldsymbol{u}=[u, v, w]$ is velocity vector, and $p$ is pressure. We consider time-varying wall-driven flow (WDF) \cite{Linot_Schmid_Taira_2024} with boundary conditions and initial conditions as:  
\begin{align}
  U(y=\pm 1, t) = \pm g_w(t), \;\;U(y,t=0)=g_w(0) y,
  \label{eq:bc}
\end{align}
where we consider exponentially decaying deceleration as
\begin{align}
\label{eq:gw_dec}
g_w(t)=\mathrm{e}^{-\kappa t},
\end{align}
and exponentially decaying acceleration as
\begin{align}
g_w(t)=1-\mathrm{e}^{-\kappa t}.
\label{eq:gw_acc}
\end{align}
In Eq. \eqref{eq:NS}, the length is normalized by the channel half-height $h$ and velocity is normalized by the maximum wall velocity $U_m$ over an infinite time, which defines the Reynolds number as $\text{Re}=U_m h / \nu$ with $\nu$ as kinematic viscosity. 

We then solve the equation for laminar base flow $\partial_t U =\partial_y^2 U/\text{Re}$ with boundary condition and initial condition in \eqref{eq:bc} to obtain the time-varying laminar base flow \cite[Section~2.1]{Linot_Schmid_Taira_2024}: 
\begin{align}
    U(y, t)=& \sum_{n=1}^{\infty} \mathrm{e}^{-a_n t}\zeta \sin (n \pi y) +\frac{\text{Re}}{6} \frac{\mathrm{~d} g_w}{\mathrm{~d} t}\left(y^3-y\right)\nonumber \\ &+g_w(t) y,
 \label{eq:laminar_solution}
\end{align}
where
\begin{subequations}
    \label{eq:zeta_Theta}
\begin{align}
\zeta:=& -\frac{2 \text{Re}(-1)^n}{(\pi n)^3}\left(\Theta_n
 +\left.\frac{\mathrm{d} g_w}{\mathrm{~d} t}\right|_{t=0}\right), 
 \\ \Theta_n:=& \int_0^t \mathrm{e}^{a_n t^{\prime}} \left.\frac{\mathrm{d}^2 g_w}{\mathrm{~d} t^2}\right|_{t=t^{\prime}} \mathrm{d} t^{\prime},\;\text{and }a_n:=(\pi n)^2/\text{Re}.   
 \label{eq:mean_flow_integral}
\end{align}
\end{subequations}

\section{\label{sec:stability}Transient Growth of Accelerating and Decelerating Flows}

This section presents how to compute the transient growth based on the SVD of the state-transition matrix following Ref. \cite{Linot_Schmid_Taira_2024}, which will be used to validate our Lyapunov-based prediction. We decompose the flow into a laminar base state $\boldsymbol{U}=[U(y,t), 0,0]^{\text{T}}$ and a perturbation $\boldsymbol{u}^{\prime}$, i.e., $\boldsymbol{u}=\boldsymbol{U}+\boldsymbol{u}^{\prime}$. We then assume that $\left|\boldsymbol{u}^{\prime}\right|\ll 1$ to neglect high-order terms, which results in linearized Navier-Stokes equations:
\begin{subequations}
\label{eq:linearNSE}
\begin{align}
\partial_t \boldsymbol{u}^{\prime} + \boldsymbol{U} {\cdot} \nabla \boldsymbol{u}^{\prime} + \boldsymbol{u}^{\prime} {\cdot} \nabla \boldsymbol{U} =& -\nabla p^{\prime} + \nabla^2 \boldsymbol{u}^{\prime}/\text{Re},\\
\nabla{\cdot} \boldsymbol{u}^{\prime}=&0. 
\end{align}
\end{subequations}

The time dependence is introduced in Eq. \eqref{eq:linearNSE} due to the time-dependence of ${U}(y,t)$. We then project Eq. \eqref{eq:linearNSE} onto the basis of the wall-normal velocity $v^{\prime}$ and the wall-normal vorticity $\omega_y^{\prime}=\partial_z u^{\prime}-\partial_x w^{\prime}$ by eliminating the pressure term using the continuity equation \cite{schimd2001stability}. We also perform the Fourier transform in streamwise $x$ and spanwise $z$ directions $
\psi^{\prime}(x, y, z, t)=\hat{\psi}(y, t) e^{\text{i}(k_x x+k_z z)}, \;\psi=v,\omega_y$, where $\text{i}=\sqrt{-1}$ is the imaginary unit. Then, we can obtain a linear time-varying system \cite{Linot_Schmid_Taira_2024} as:
\begin{equation}
\partial_t \boldsymbol{q}= \boldsymbol{L}(t)\boldsymbol{q}  ,
\label{eq:discrtizedformlinearsystem}
\end{equation}
where 
\begin{align}
    \boldsymbol{q}:=\begin{bmatrix}
        \hat{v}\\
        \hat{\omega}_y
    \end{bmatrix},\;\;\;\boldsymbol{L}(t):=\left[\begin{array}{cc}
\mathscr{L}_{\mathrm{os}} & 0 \\
\mathscr{L}_{\mathrm{c}} & \mathscr{L}_{\mathrm{sq}}
\end{array}\right]
\end{align}
with
\begin{subequations}
    \begin{align}
\mathscr{L}_{\mathrm{os}}:=&(\partial_y^2-k^2)^{-1}[(\partial_y^2
-k^2)^2/Re \nonumber\\
&-\mathrm{i} k_x U(\partial_y^2-k^2)+ \mathrm{i} k_x \partial_y^2 U],\\
\mathscr{L}_{\mathrm{c}}:=& -\text{i} k_z \partial_y U,\\
\mathscr{L}_{\mathrm{sq}}:=& - \text{i} k_x U + \left(\partial_y^2-k^2\right)/Re,
\label{eq28}
\end{align}
\end{subequations}
and $k^2=k_x^2+k_z^2$. The boundary conditions are $
\hat{v}(y= \pm 1)=\partial_y\hat{v}(y= \pm 1)=\hat{\omega}_y(y= \pm 1)=0$. 

Here, we are interested in transient growth defined as:
\begin{align}
    \label{eq:G_def_energy}G\left(t;k_x,k_z,t_0,Re,\kappa\right):=\max _{\boldsymbol{q}\left(t_0\right) \neq \boldsymbol{0}} \frac{\|\boldsymbol{q}(t)\|_E^2}{\left\|\boldsymbol{q}\left(t_0\right)\right\|_E^2},
\end{align}
where the energy norm is defined as:
\begin{dmath}
\|\boldsymbol{q}\|_E^2:=\int_{-1}^1\left(|\partial_y \hat{v}|^2+k^2|\hat{v}|^2+\left|\hat{\omega}_y\right|^2\right) \mathrm{d} y. \label{eq:energy_norm}
\end{dmath}
As shown by \cite{gustavsson1986excitation}, dividing the energy norm in \eqref{eq:energy_norm} by $2k^2$ and integrating over wavenumbers yields the total kinetic energy. The quantity $G(t)$ represents the maximal energy gain achievable by any initial perturbation $\boldsymbol{q}(t_0)$ over $[t_0,t]$. We omit the explicit dependence on parameters $k_x$, $k_z$, $t_0$, $\text{Re}$, and $\kappa$ for $G(t)$ to simplify our notation in what follows. 

To transform the energy norm to the vector 2-norm such that $\|\boldsymbol{q}\|_E^2=\|\boldsymbol{x}\|_2^2$, we use a change of variable 
\begin{align}
\label{eq:change_variable_q2x}
    \boldsymbol{x} = \boldsymbol{S} \boldsymbol{q}
\end{align}
based on \cite[Appendix B]{Linot_Schmid_Taira_2024} leading to a state-space model
\begin{equation}
\partial_t \boldsymbol{x}= \boldsymbol{A}(t)\boldsymbol{x},
\label{eq:discrtizedformlinearsystem}
\end{equation}
where $\boldsymbol{A}(t)=\boldsymbol{S}\boldsymbol{L}(t)\boldsymbol{S}^{-1}$. This linear time-varying system in Eq. \eqref{eq:discrtizedformlinearsystem} has the solution of the form
\begin{equation}
\boldsymbol{x}(t)=\boldsymbol{\Phi}(t,t_0) \boldsymbol{x}(t_0),
\label{eq31}
\end{equation}
where $\boldsymbol{\Phi}(t,t_0)$ is the state-transition matrix that satisfies
\begin{equation}
\partial_t  \boldsymbol{\Phi}(t,t_0)=\boldsymbol{A}(t) \boldsymbol{\Phi}(t,t_0),\quad \boldsymbol{\Phi}(t_0,t_0)=\boldsymbol{I}
\label{eq32}
\end{equation}
with $\boldsymbol{I}$ as the identity matrix. We then numerically approximate $\boldsymbol{\Phi}(t,t_0)$ at $t=t_0+n\Delta t$ as 
\begin{subequations}
\label{eq:state_transition_matrix}
    \begin{align}
&\boldsymbol{\Phi}(t_0+n\Delta t,t_0)=\prod_{j=1}^n e^{\boldsymbol{A}(t_0+j \Delta t) \Delta t}\\
=&e^{\boldsymbol{A}(t_0+n\Delta t) \Delta t}\cdots e^{\boldsymbol{A}(t_0+2\Delta t) \Delta t}e^{\boldsymbol{A}(t_0+\Delta t) \Delta t}
\end{align}
\end{subequations}
with a time step $\Delta t$. We use this $\boldsymbol{\Phi}(t,t_0)$ and the relation $\|\boldsymbol{q}\|_E^2=\|\boldsymbol{x}\|_2^2$ to determine the transient growth in \eqref{eq:G_def_energy} as: 
\begin{subequations}
\label{eq:G_svd_full}
\begin{align}
G(t)=&\max _{\boldsymbol{x}\left(t_0\right) \neq \boldsymbol{0}}\frac{\|\boldsymbol{x}(t)\|_2^2}{\|\boldsymbol{x}(t_0)\|_2^2}\label{eq:G_definition} \\
=&\max _{\boldsymbol{x}\left(t_0\right) \neq \boldsymbol{0}}\frac{\|\boldsymbol{\Phi}(t,t_0)\boldsymbol{x}(t_0)\|_2^2}{\|\boldsymbol{x}(t_0)\|_2^2}=\|\boldsymbol{\Phi}(t,t_0)\|_2^2.
\label{eq:G_svd}
\end{align}
\end{subequations}
The last term in Eq. \eqref{eq:G_svd} is based on the matrix-induced norm that can be computed by SVD of $\boldsymbol{\Phi}(t,t_0)$, which will be used to validate the Lyapunov-based upper bound.

\section{Upper Bound of Transient Growth Using Lyapunov Method}
\label{sec:upper_bounds}
This section presents our core contribution: a Lyapunov-based approach for certifying the uniform stability, bounding maximal transient energy growth, and constructing an invariant set for time-varying shear flows, which will be formulated in Theorem \ref{thm:LMI}. Similar LMI for time-invariant systems has been used in, e.g., Refs. \cite{kalur2021nonlinear,whidborne2007minimization}. We start by introducing the uniform stability of time-varying systems.  
\newtheorem{definition}{Definition} %
\begin{definition}[Definition 4.4 in \cite{khalil2002nonlinear}]
The equilibrium point $\boldsymbol{x}=0$ of \eqref{eq:discrtizedformlinearsystem} is uniformly stable if, for each $\epsilon>0$, there is $\delta=\delta(\epsilon)>0$, independent of $t_0$, such that $\|\boldsymbol{x}(t_0)\|<\delta \Rightarrow \|\boldsymbol{x}(t)\|<\epsilon,\;\forall t\geq t_0>0$. 
\end{definition}

\newtheorem{theorem}{Theorem}
\begin{theorem}\label{thm:LMI}
Given a linear time-varying system in \eqref{eq:discrtizedformlinearsystem}. If we can find a continuously differentiable Hermitian matrix $\boldsymbol{P}(t)$ and $\overline{G}>0$ by solving 
\begin{subequations}
\label{eq:LMI}
    \begin{align}
&\text{min}& & \overline{G} \label{eq:LMI_optimal} \\
&\text{subject to}& & \boldsymbol{I} \preceq \boldsymbol{P}(t) \preceq \overline{G} \boldsymbol{I},\label{eq:LMI_constraints_P} \\
&& &  \dot{\boldsymbol{P}}(t)+ \boldsymbol{A}^*(t)\boldsymbol{P}(t) + \boldsymbol{P}(t)\boldsymbol{A}(t)  \preceq 0,
\label{eq:LMI_constraints}
\end{align}
\end{subequations}
where $(\cdot)\preceq0$ means negative semi-definiteness. Then, the equilibrium $\boldsymbol{x}=\boldsymbol{0}$ of system \eqref{eq:discrtizedformlinearsystem} is uniformly stable and the transient energy growth $G(t)$ is bounded by:
\begin{equation}
G(t) \leq \overline{G}.
\label{eq:bound}
\end{equation}
\end{theorem}

\begin{proof}
Consider a Lyapunov function $V(\boldsymbol{x},t)=\boldsymbol{x}^*\boldsymbol{P}(t)\boldsymbol{x}$, we then compute $\dot{V}$ as:
\begin{subequations}
\label{eq49}
\begin{align}
\dot{V} &= \boldsymbol{x}^* \dot{\boldsymbol{P}} \boldsymbol{x} + \dot{\boldsymbol{x}}^* \boldsymbol{P} \boldsymbol{x}  + \boldsymbol{x}^* \boldsymbol{P} \dot{\boldsymbol{x}} \\
&= \boldsymbol{x}^* \dot{\boldsymbol{P}} \boldsymbol{x}+ \boldsymbol{x}^* \boldsymbol{A}^* \boldsymbol{P} \boldsymbol{x}  + \boldsymbol{x}^* \boldsymbol{P} \boldsymbol{A} \boldsymbol{x} \\
&= \boldsymbol{x}^* \left(\dot{\boldsymbol{P}} + \boldsymbol{A}^* \boldsymbol{P} + \boldsymbol{P} \boldsymbol{A} \right) \boldsymbol{x}.
\end{align}
\end{subequations}
If $\dot{\boldsymbol{P}}+ \boldsymbol{A}^* \boldsymbol{P} + \boldsymbol{P} \boldsymbol{A} \preceq 0$ based on Eq. \eqref{eq:LMI_constraints}, then $\dot{V} \leq 0,\;\forall\, t$, which implies that $V(\boldsymbol{x}(t),t) \leq V(\boldsymbol{x}(t_0),t_0),\;\forall\, t \geq t_0$. Then, all requirements in Theorem 4.8 of Ref. \cite{khalil2002nonlinear} are satisfied, which proves that $\boldsymbol{x}=\boldsymbol{0}$ of \eqref{eq:discrtizedformlinearsystem} is uniformly stable. Then, we have 
\begin{align}
& \lambda_{\min}[\boldsymbol{P}(t)] \boldsymbol{x}^*(t) \boldsymbol{x}(t) \leq \boldsymbol{x}^*(t) \boldsymbol{P}(t) \boldsymbol{x}(t) = V(\boldsymbol{x}(t),t)\nonumber\\
\leq& V(\boldsymbol{x}(t_0),t_0)
  =\boldsymbol{x}(t_0)^* \boldsymbol{P}(t_0) \boldsymbol{x}(t_0)\nonumber \\
  \leq& \lambda_{\max}[\boldsymbol{P}(t_0)] \boldsymbol{x}^*(t_0) \boldsymbol{x}(t_0),
  \label{eq50}
\end{align}
where $\lambda_{\min}[\cdot]$ and $\lambda_{\max}[\cdot]$ are respectively the minimal and maximal eigenvalues of the argument. Thus, we have:
\begin{equation}
G(t)=\max _{\boldsymbol{x}\left(t_0\right) \neq \boldsymbol{0}}\frac{\boldsymbol{x}^*(t) \boldsymbol{x}(t)}{\boldsymbol{x}^*(t_0) \boldsymbol{x}(t_0)} \leq \frac{\lambda_{\max}[\boldsymbol{P}(t_0)]}{\lambda_{\min}[\boldsymbol{P}(t)]} \leq \overline{G}.
\label{eq52}
\end{equation}
\end{proof}

\todo{Theorem \ref{thm:LMI} differs from its time-invariant counterpart (e.g., \cite{kalur2021nonlinear,whidborne2007minimization}) by using a time-varying $\boldsymbol{P}(t)$ with the $\dot{\boldsymbol{P}}(t)$ term in \eqref{eq:LMI_constraints}, leading to a time-varying Lyapunov function $V(\boldsymbol{x},t)=\boldsymbol{x}^*\boldsymbol{P}(t)\boldsymbol{x}$ to reduce conservatism compared with the Lyapunov function candidate $V(\boldsymbol{x})=\boldsymbol{x}^*\boldsymbol{P}\boldsymbol{x}$.} Without the $\dot{\boldsymbol{P}}(t)$ term in Eq. \eqref{eq:LMI_constraints}, the resulting upper bound $\overline{G}$ will be extremely conservative; i.e., $\overline{G}\gg G(t)$. We discretize time as $t_i$ with a fixed time step $\Delta t=t_{i+1}-t_i$ and employ a forward Euler scheme to approximate $\dot{\boldsymbol{P}}(t_i)\approx \frac{\boldsymbol{P}(t_{i+1}) - \boldsymbol{P}(t_i)}{\Delta t}$, which modifies Eq. \eqref{eq:LMI} as
\begin{subequations}
    \label{eq:discrete_LMI}
\begin{align}
&\text{min }  \overline{G}\\
&\text{subject to } \boldsymbol{I} \preceq \boldsymbol{P}(t_i) \preceq \overline{G} \boldsymbol{I}, \;\;\forall i = 0, 1, 2, \cdots, n,  \\
&\frac{\boldsymbol{P}(t_{i+1}) - \boldsymbol{P}(t_i) }{\Delta t}+  \boldsymbol{A}^*(t_i)\boldsymbol{P}(t_i) + \boldsymbol{P}(t_i)\boldsymbol{A}(t_i) \preceq 0. 
\end{align}
\end{subequations}

We then implement LMI stated in \eqref{eq:discrete_LMI} using YALMIP version R20230622 \cite{Lofberg2004} within MATLAB R2022a and the semidefinite programming solver Mosek 10.2 \cite{Mosek2024}. To validate our upper bound prediction, we also compute $G(t)$ for WDF by approximating $\boldsymbol{\Phi}(t,t_0)$ using Eq. \eqref{eq:state_transition_matrix}. We discretize the linear operator $\boldsymbol{L}(t)$ in the wall-normal direction $y$ using \todo{the central difference} scheme with second-order accuracy. We use grid points $M=32$ (including points at the boundary), which determines the dimension of our linear time-varying systems, i.e., $\boldsymbol{x}\in \mathbb{C}^{2(M-2)\times 1}$ and $\boldsymbol{A}(t)\in \mathbb{C}^{2(M-2)\times 2(M-2)}$. We use the same discretization scheme in the wall-normal direction with $M=32$ and select a fixed time step of $\Delta t=1$ for computing both the upper bound $\overline{G}$ in Eq. \eqref{eq:discrete_LMI} and transient growth $G(t)$ in Eq. \eqref{eq:G_svd} for a direct comparison.  \todo{Although higher-order finite-difference schemes can be employed, the grid-refinement study in Section \ref{sec:results} indicates that the central difference scheme already yields converged results.} We approximate the base flow with $n=100$ basis functions of $\sin(n \pi y)$ in Eq. \eqref{eq:laminar_solution}. In this paper, we consider accelerating and decelerating WDF at $\operatorname{Re}=500$ and $\kappa=0.1$, which allows us a direct comparison with Ref. \cite{Linot_Schmid_Taira_2024}. 

\section{Results and Discussions}
\label{sec:results}
\begin{figure}[!htbp]
    \centering
    \includegraphics[width=0.46\textwidth]{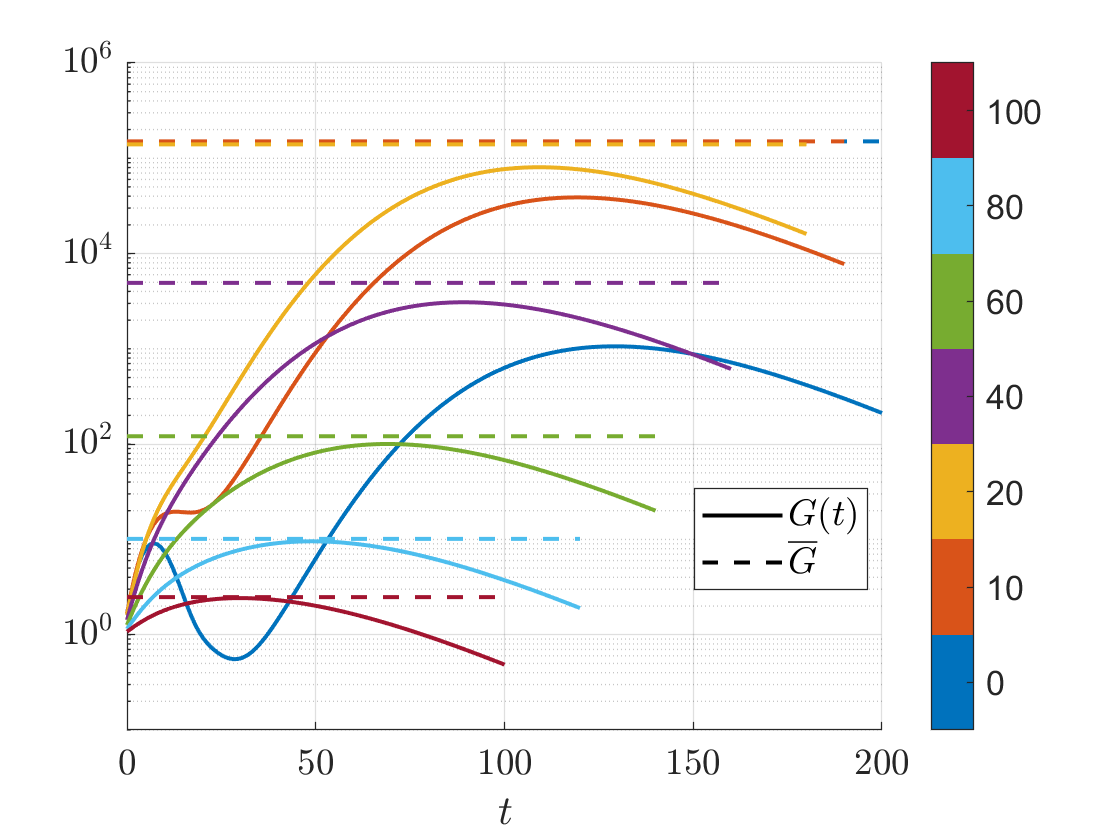}
    \caption{The transient growth $G(t)$ and its upper bound $\overline{G}$ for decelerating WDF at $\text{Re} = 500$, $\kappa = 0.1$, and $[k_x, k_z] = [1.2, 0]$. Each color corresponds to a different initial time at $t_0 = [0, 10, 20, 40, 60, 80, 100]$.}
    \label{dec_WDF_1.2_0}
\end{figure}

Figure \ref{dec_WDF_1.2_0} shows the transient growth $G(t)$ and its upper bound $\overline{G}$ for decelerating WDF ($g_w(t)$ as Eq. \eqref{eq:gw_dec}) associated with streamwise dependent and spanwise invariant perturbations (i.e., $[k_x, k_z] = [1.2, 0]$) and different initial time $t_0$. Here, for all $t_0$, the upper bound obtained from the Lyapunov method $\overline{G}$ is larger than and close to the maximal transient growth $\underset{t}{\max}\, G(t)$. The maximal transient growth $\underset{t}{\max}\,G(t)$ increases for $t_0\in [0,20]$ and then decreases at $t_0\in [20,100]$, while the upper bound $\overline{G}$ remains the same for $t_0\in [0,20]$ (overlapped to the orange horizontal line of $t_0=20$). \todo{Reducing $\Delta t$ in computing $\overline{G}$ at $t_0=0$ leads to similar upper bound $\overline{G}$, which suggests that the difference between $\overline{G}$ and $\underset{t}{\max}\,G(t)$ at small $t_0$ is likely an intrinsic feature of the Lyapunov method.} The upper bound $\overline{G}$ then decreases within $t_0\in [20,100]$ and closely follows the maximal transient growth $\underset{t}{\max}\,G(t)$. 

\begin{figure}[!htbp]
    \centering
    \includegraphics[width=0.49\textwidth]{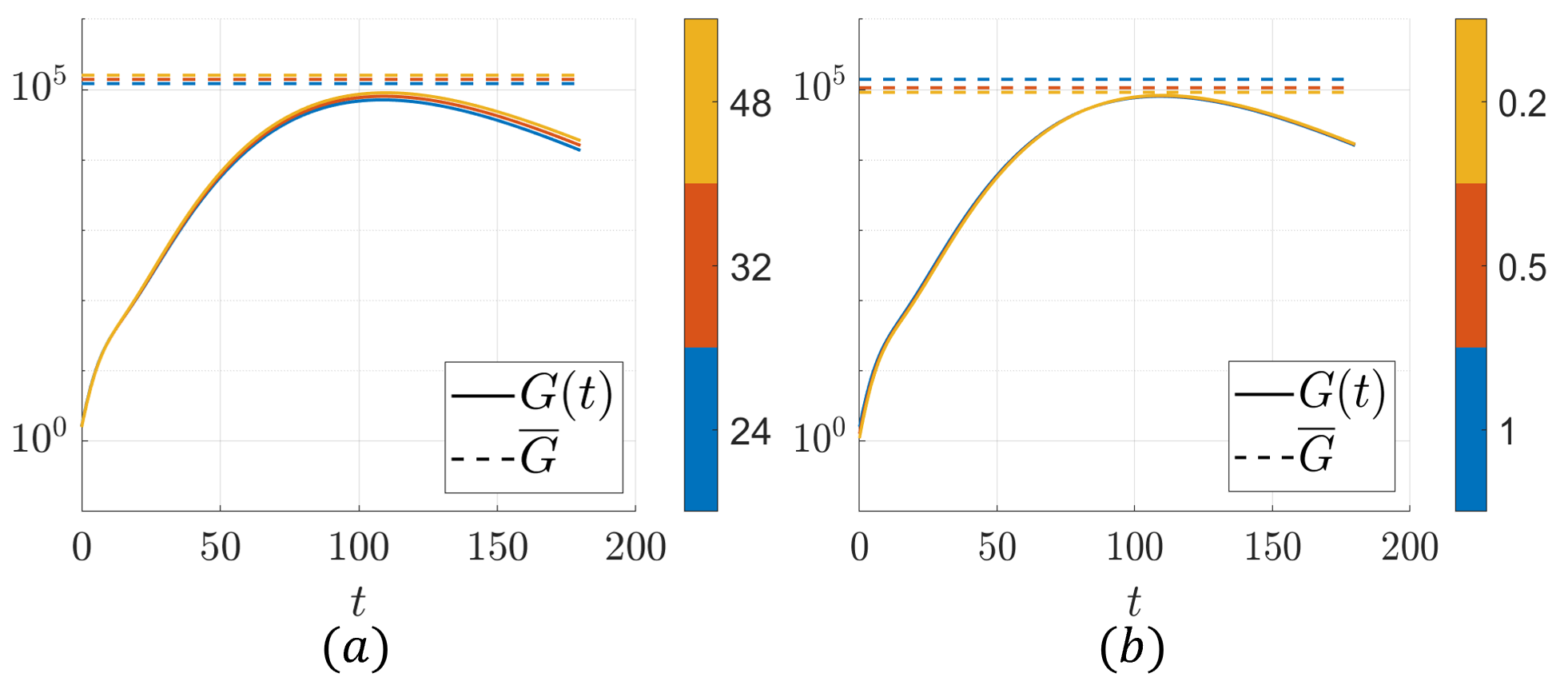}
    \caption{The transient growth $G(t)$ and its upper bound $\overline{G}$ for decelerating WDF at $\text{Re} = 500$, $\kappa = 0.1$, $t_0 = 20$, and $[k_x, k_z] = [1.2, 0]$. Panel (a) shows results with $\Delta t=1$ and $M=$24, 32, and 48 (colorbar). Panel (b) shows results with $M=32$ and $\Delta t = 1$, 0.5, and 0.2 (colorbar).}
\label{Different_grid_time_step}
\end{figure}

Figure \ref{Different_grid_time_step} shows how the wall-normal grid point number $M$ and the time step $\Delta t$ affect the transient growth $G(t)$ and its upper bound $\overline{G}$ from the Lyapunov method. In Fig.~\ref{Different_grid_time_step}(a), increasing $M$ from 24 to 48 leads to a slightly higher peak in transient growth $G(t)$ and upper bound $\overline{G}$, but the influence of the grid point number $M$ is very small. Figure \ref {Different_grid_time_step}(b) shows that decreasing the time step $\Delta t$ will lead to a tighter upper bound $\overline{G}$, i.e., $\overline{G}$ is closer to $\underset{t}{\max}\,G(t)$, while changing different $\Delta t$ shows negligible influence on $G(t)$. \todo{Table \ref{tab_comp_cost} shows the computational time and memory requirements for computing $\overline{G}$ in Fig. \ref{Different_grid_time_step} associated with different grid resolutions $M$ and time steps $\Delta t$, highlighting the significant increase in computational cost required to obtain a tighter upper bound $\overline{G}$, which is in general more computationally expensive than computing $G(t)$ based on SVD in \eqref{eq:G_svd_full}.}

\begin{table}[!htbp]
\caption{Computational cost for obtaining $\overline{G}$ in Fig. \ref{Different_grid_time_step} using LMI in \eqref{eq:discrete_LMI}.}
\begin{center}
\begin{tabular}{|c|c|c|c|c|}
\hline
\textbf{Panel}&\multicolumn{4}{|c|}{\textbf{Simulation Parameters and Costs}} \\
\cline{2-5} 
\textbf{Case} & $M$& $\Delta t$& Time (s) & Memory (GB) \\
\hline
Panel (a) & 24 & 1 & 4861.81 & 21.32 \\
\cline{2-5}
 & 32 & 1 & 13354.93 & 67.68 \\
\cline{2-5}
& 48 & 1 & 272660.37 & 539.58 \\
\hline
Panel (b) & 32 & 1 & 13354.93 & 67.68 \\
 \cline{2-5}
 & 32 & 0.5 & 41247.93 & 189.31 \\
\cline{2-5}
& 32 & 0.2 & 112825.82 & 367.37 \\
\hline
\end{tabular}
\label{tab_comp_cost}
\end{center}
\end{table}

\begin{figure}[!htbp]
    \centering
    \includegraphics[width=0.46\textwidth]{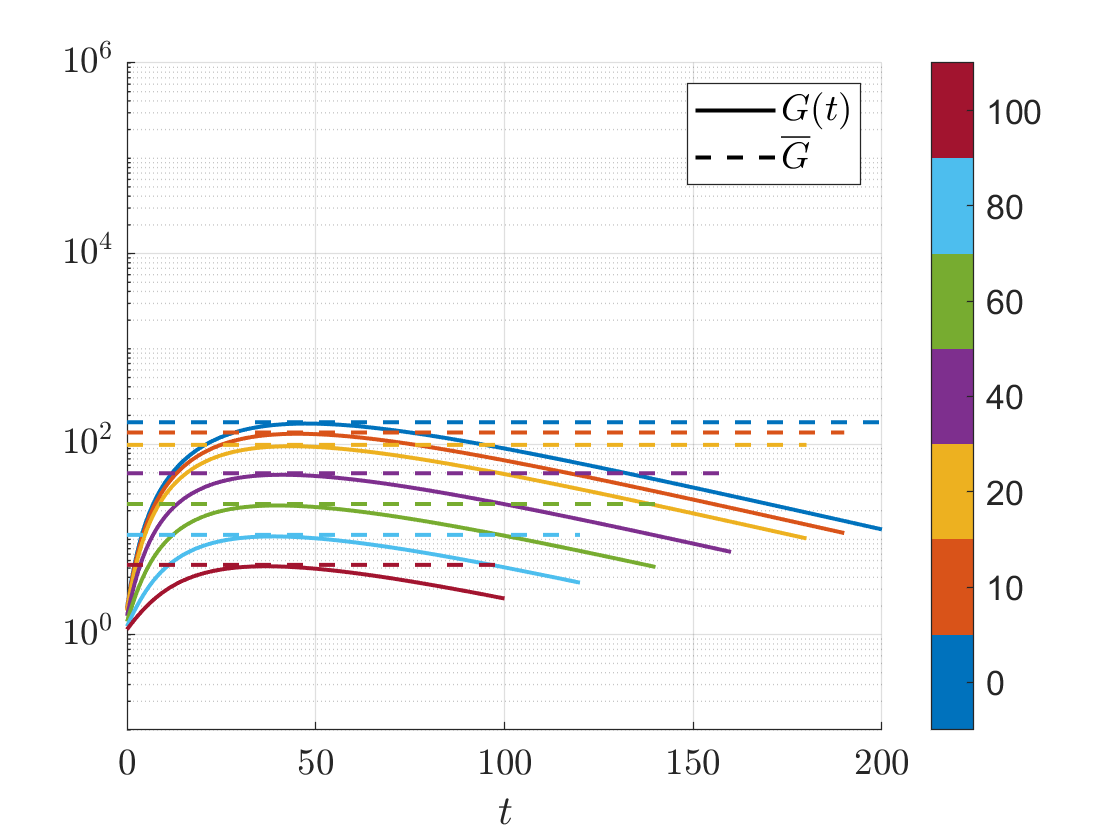}
    \caption{The transient growth $G(t)$ and its upper bound $\overline{G}$ for decelerating WDF at $\text{Re} = 500$, $\kappa = 0.1$, and $[k_x, k_z] = [0, 1.6]$. Each color corresponds to a different initial time at $t_0 = [0, 10, 20, 40, 60, 80, 100]$. }
    \label{Dec_WDF_0_1.6}
\end{figure}

\todo{We then consider a different Fourier mode representing streamwise invariant and spanwise dependent perturbations ($[k_x, k_z] = [0, 1.6]$), and} Fig. \ref{Dec_WDF_0_1.6} shows the corresponding transient growth $G(t)$ and its upper bound $\overline{G}$ of decelerating WDF. Here, we can also find that the upper bound from the Lyapunov method $\overline{G}$ is a valid upper bound, i.e., $G(t)\leq\overline{G}$. Moreover, $\overline{G}$ is very close to $\underset{t}{\max}\,G(t)$ for all $t_0$. 

\begin{figure}[!htbp]
    \centering
    \includegraphics[width=0.46\textwidth]{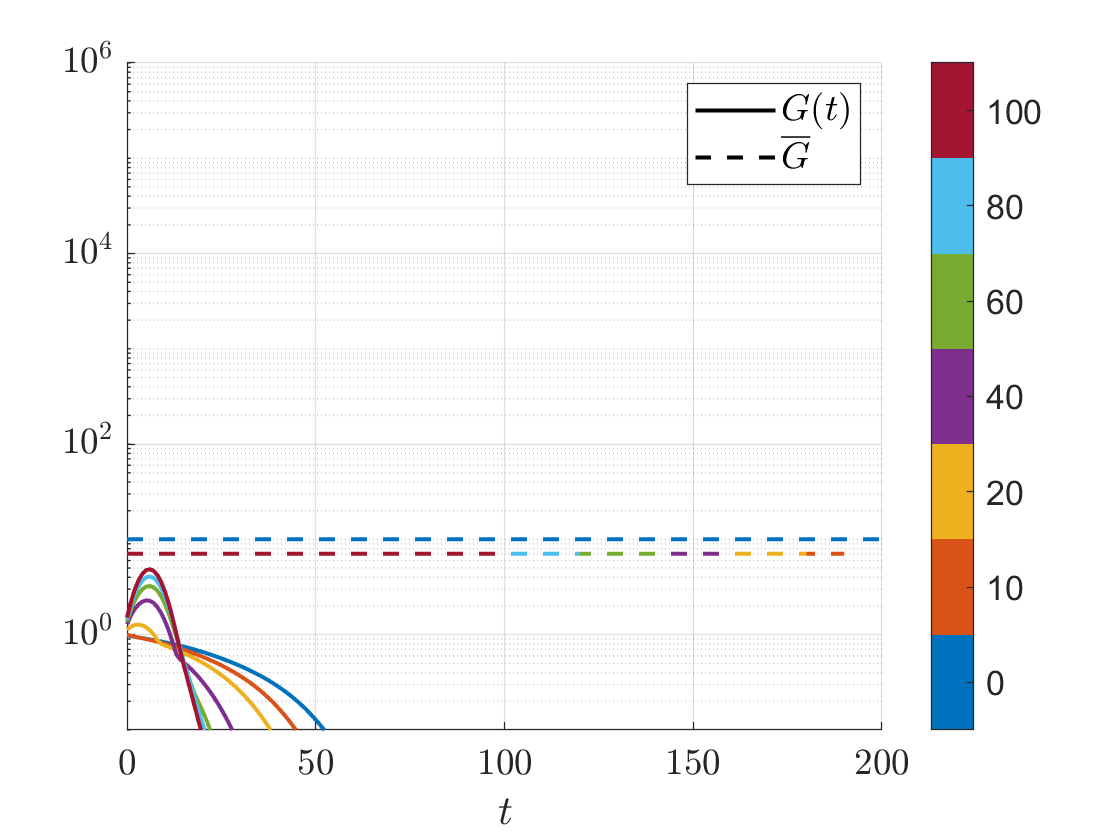}
    \caption{The transient growth $G(t)$ and its upper bound $\overline{G}$ for accelerating WDF at $\text{Re} = 500$, $\kappa = 0.1$, and $[k_x, k_z] = [1.2, 0]$. Each color corresponds to a different initial time at $t_0 = [0, 10, 20, 40, 60, 80, 100]$. }
    \label{Acc_WDF_1.2_0}
\end{figure}

\begin{figure}[!htbp]
    \centering
    \includegraphics[width=0.46\textwidth]{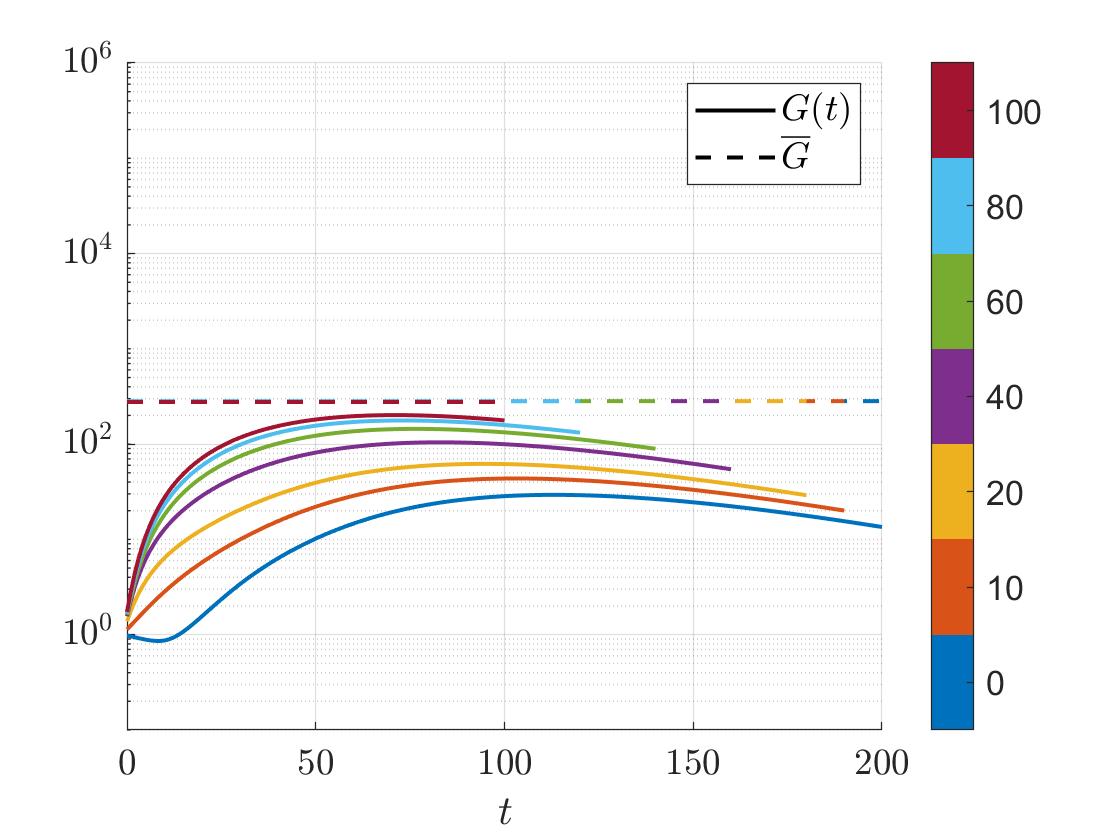}
    \caption{The transient growth $G(t)$ and its upper bound $\overline{G}$ for accelerating WDF at $\text{Re} = 500$, $\kappa = 0.1$, and $[k_x, k_z] = [0, 1.6]$. Each color corresponds to a different initial time at $t_0 = [0, 10, 20, 40, 60, 80, 100]$.}
    \label{Acc_WDF_0_1.6}
\end{figure}

Figures \ref{Acc_WDF_1.2_0}-\ref{Acc_WDF_0_1.6} then consider accelerating WDF ($g_w(t)$ in Eq. \eqref{eq:gw_acc}) with respectively $[k_x, k_z] = [1.2, 0]$ in Fig. \ref{Acc_WDF_1.2_0} and $[k_x, k_z] = [0, 1.6]$ in Fig. \ref{Acc_WDF_0_1.6}, which are the same wavenumber chosen in Figs. \ref{dec_WDF_1.2_0} and \ref{Dec_WDF_0_1.6} \cite{Linot_Schmid_Taira_2024}. Here, the upper bounds $\overline{G}$ associated with different $t_0$ overlap to two horizontal lines in Fig. \ref{Acc_WDF_1.2_0} and overlap to one horizontal line in Fig. \ref{Acc_WDF_0_1.6}, which is different from the upper bound $\overline{G}$ obtained in decelerating flows (Figs. \ref{dec_WDF_1.2_0} and \ref{Dec_WDF_0_1.6}) that closely follow the maximal transient growth. For streamwise varying mode ($k_x\neq 0$), transient growth in accelerating flows (Fig. \ref{Acc_WDF_1.2_0}) is much smaller than that in decelerating flows (Fig. \ref{dec_WDF_1.2_0}), \todo{which is consistent with the flow physics that acceleration will stabilize the flow while deceleration will destabilize the flow \cite{Linot_Schmid_Taira_2024,linot2025extractingdominantdynamicsunsteady}. The destabilizing effects of decelerating flows may also lead to the upper bound $\overline{G}$ closely following the maximum of transient growth $G(t)$ in decelerating flows (Figs. \ref{dec_WDF_1.2_0} and \ref{Dec_WDF_0_1.6}).} For spanwise varying mode ($k_z\neq 0$), the transient growth $G(t)$ will decrease as increasing initial time $t_0$ in decelerating flow (Fig. \ref{Dec_WDF_0_1.6}), while it will increase as increasing $t_0$ in accelerating flow (Fig. \ref{Acc_WDF_0_1.6}). For all results in Figs. \ref{dec_WDF_1.2_0}--\ref{Acc_WDF_0_1.6}, we successfully find a Lyapunov function $\boldsymbol{P}(t)$, which also certify the uniform stability of the equilibrium $\boldsymbol{x}=\boldsymbol{0}$ for linear time-varying system in \eqref{eq:discrtizedformlinearsystem}.

\begin{figure}[!htbp]
    \centering
    \includegraphics[width=0.5\textwidth]{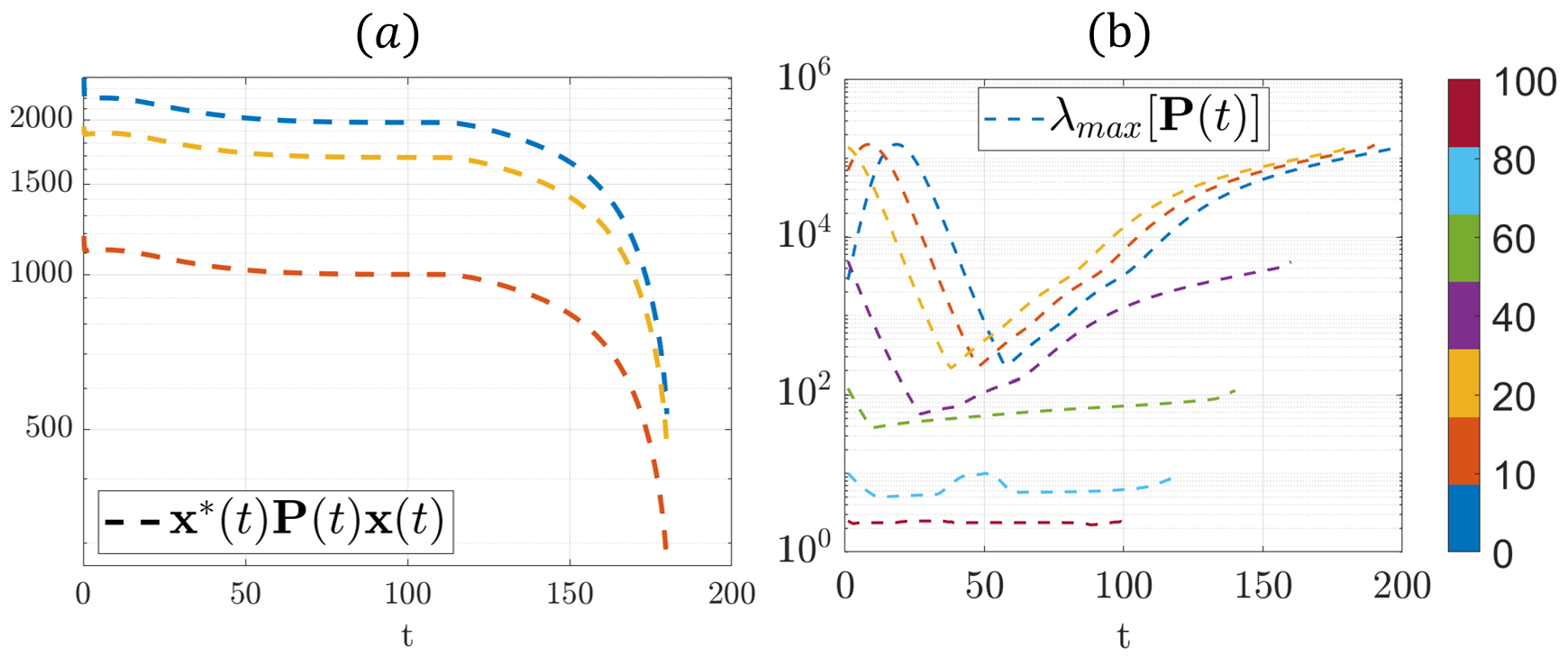}
    \caption{(a) Evolution of the Lyapunov function $V(\boldsymbol{x},t)=\boldsymbol{x}^{\ast}(t) \boldsymbol{P}(t) \boldsymbol{x}(t)$ ($\dashed$) with unit norm random initial conditions, where three different colors correspond to different random initial conditions. Results are associated with decelerating WDF at $\text{Re} = 500$, $\kappa = 0.1$, $\Delta t = 0.2$, $t_0 = 20$, and $[k_x, k_z] = [1.2, 0]$. (b) Maximal eigenvalues of $\boldsymbol{P}(t)$ matrix over $t$ for decelerating WDF at $\text{Re} = 500$, $\kappa = 0.1$, and $[k_x, k_z] = [1.2, 0]$. Each color corresponds to a different initial time at $t_0 = [0, 10, 20, 40, 60, 80, 100]$. }
    \label{Lyap_and_bound_max_eigen}
\end{figure}

We then show that our Lyapunov function can be used to construct an invariant set based on $V(\boldsymbol{x}(t),t)\leq V(\boldsymbol{x}(t_0),t_0)$ in Eq. \eqref{eq50}, which leads to a bound on solution trajectory as $\boldsymbol{x}^*(t)\boldsymbol{P}(t)\boldsymbol{x}(t)\leq \boldsymbol{x}^*(t_0)\boldsymbol{P}(t_0)\boldsymbol{x}(t_0)$. Here, we focus on decelerating WDF with $[k_x,k_z]=[1.2,0]$ associated with Fig. \ref{dec_WDF_1.2_0}, as this parameter regime shows large transient growth. Figure \ref{Lyap_and_bound_max_eigen}(a) shows that $V(\boldsymbol{x},t)=\boldsymbol{x}^*(t)\boldsymbol{P}(t)\boldsymbol{x}(t)$ is not increasing over time, consistent with our LMI in Theorem \ref{thm:LMI} such that $\dot{V}\leq 0$. Here, we set the initial condition of unit norm (i.e., $\|\boldsymbol{x}(t_0)\| = 1$) and obtain the solution $\boldsymbol{x}(t)$ by Eq. \eqref{eq31} with the state-transition matrix obtained in Eq. \eqref{eq:state_transition_matrix}. This invariant set $\boldsymbol{x}^*(t)\boldsymbol{P}(t)\boldsymbol{x}(t)\leq \boldsymbol{x}^*(t_0)\boldsymbol{P}(t_0)\boldsymbol{x}(t_0)$ allows us to characterize the solution trajectory in more detail than transient growth analysis $\boldsymbol{x}^*(t)\boldsymbol{x}(t)\leq G(t)\boldsymbol{x}^*(t_0)\boldsymbol{x}(t_0)$. 

\todo{Figure \ref{Lyap_and_bound_max_eigen}(b) shows how the maximal eigenvalues of the time‐dependent Lyapunov matrix $\boldsymbol{P}(t)$ evolve for the decelerating WDF, which can reflect the shape of the invariant set constructed by the Lyapunov function. The maximal eigenvalue $\lambda_{\max}[\boldsymbol{P}(t)]$ peaks initially and then decays as the shear weakens, and finally rises again. Here, the time reaching maximal $\lambda_{\max}[\boldsymbol{P}(t)]$ is different from the time leading to maximal transient growth $G(t)$ in Fig. \ref{dec_WDF_1.2_0}. The minimal eigenvalue remains $\mathcal{O}(1)$ for all $t$ with $\lambda_{\min}[\boldsymbol{P}(t)]\geq 1$ (not shown), consistent with the constraint in Eq. \eqref{eq:LMI_constraints_P}.}

\begin{figure}[!htbp]
    \centering
    \includegraphics[width=0.47\textwidth]{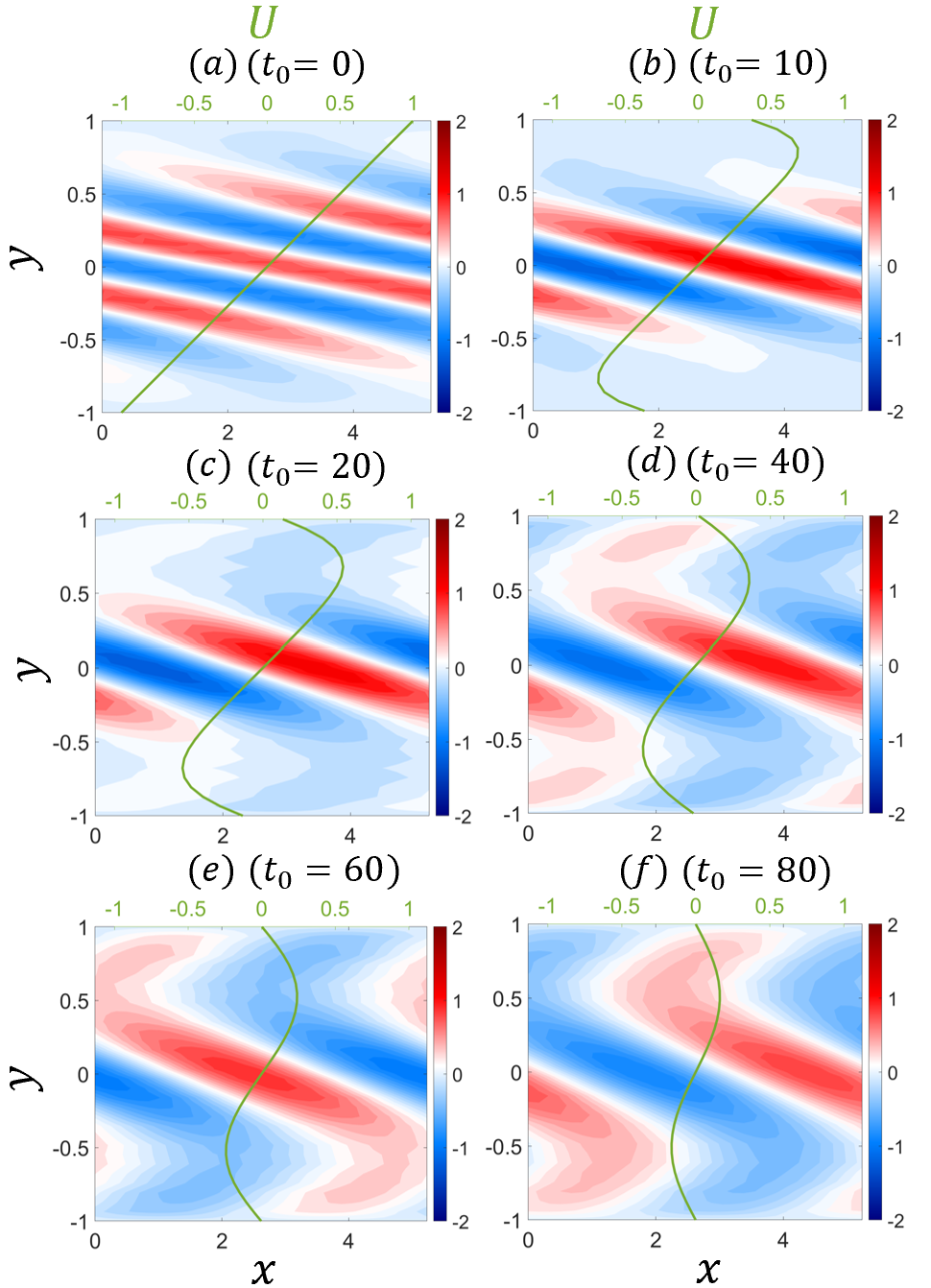}
    \caption{Primary streamwise velocity modes $u$ from the eigenvector corresponding to $\lambda_{\text{max}}[\boldsymbol{P}(t_0)]$ for decelerating WDF at $\text{Re} = 500$, $\kappa = 0.1$, and $[k_x, k_z] = [1.2, 0]$. Panels (a)-(f) are associated with $t_0 = [0, 10, 20, 40, 60, 80]$, and the green line is the laminar base flow $U(y,t_0)$. 
    }
    \label{velocity_mode}
\end{figure}

Figure \ref{velocity_mode} shows the primary streamwise velocity modes $u$ by computing the eigenvector $\boldsymbol{x}$ associated with $\lambda_{\text{max}}[\boldsymbol{P}(t_0)]$, which will maximize $V(t_0)$. We convert this eigenvector back to state variable $\boldsymbol{q}$ based on Eq. \eqref{eq:change_variable_q2x} and then obtain streamwise velocity via $\hat{u}=\frac{1}{k_x^2+k_z^2}\begin{bmatrix}     \text{i}k_x\partial_y & -\text{i}k_z\end{bmatrix}\boldsymbol{q}$. Such Fourier-space amplitude of streamwise velocity $\hat{u}$ is then converted to physical space through $
u(x, y, z, t)= \Re\left[\hat{u}\left(y,t; k_x, k_z\right) e^{\mathrm{i}\left(k_x x+k_z z\right)}\right]$, where $\Re[\cdot]$ means the real part of the argument. In Fig. \ref{velocity_mode}, we also superpose the laminar base flow $U(y,t_0)$ (green lines) based on Eq. \eqref{eq:laminar_solution}. At $t_0=0$ and 10, the primary streamwise velocity mode (contour) is inclined upstream, in the opposite direction of the laminar base flow profile (green line). Such a feature persists as $t_0$ increases, and near-wall structures are inclined in the opposite direction of those in the channel center when $t_0\geq 20$, which is due to the difference in the laminar base flow in these regions. The primary mode shifts gradually from the channel core toward the walls as deceleration proceeds, tracking the change in the laminar profile (green line). This upstream‐inclination and subsequent realignment of the primary mode is a signature of the Orr mechanism \cite{orr1907stability}, which is qualitatively similar to linear optimal perturbations identified from SVD of state-transition matrix \cite{Linot_Schmid_Taira_2024}.

\section{Conclusions}
\label{sec:conclusions}

We employ a Lyapunov‐based approach to certify the uniform stability and obtain an upper bound on transient energy growth in accelerating and decelerating wall-driven flows. By formulating the linearized Navier–Stokes equations as a time-varying system and constructing time-dependent Lyapunov functions, we derive the LMI whose solutions yield an upper bound that closely tracks transient growth obtained by singular value decomposition of the state-transition matrix. Our Lyapunov approach captures the essential flow physics, showing that decelerating flows experience substantially larger transient growth than accelerating flows, primarily due to the Orr mechanism. The Lyapunov function certifies uniform stability and provides an invariant set to effectively bound solution trajectories. In future work, we plan to \todo{employ this framework to explore wider parameter regimes ($k_x$ and $k_z$) and} extend this Lyapunov approach to input-output analysis and nonlinear analysis of \todo{stable and unstable time-varying shear flows}, see e.g., \cite{kochnev2025stability,wei2025nonlinear}. 

\bibliographystyle{ieeetr}
\bibliography{main}

@article{brunton2015closed,
  title={Closed-loop turbulence control: Progress and challenges},
  author={Brunton, Steven L and Noack, Bernd R},
  journal={Appl. Mech. Rev.},
  volume={67},
  number={5},
  pages={050801},
  year={2015},
  publisher={American Society of Mechanical Engineers}
}

@article{orr1907stability,
  title={The stability or instability of the steady motions of a perfect liquid and of a viscous liquid. Part {II}: A viscous liquid},
  author={Orr, William M'F},
  journal={{Proc. R. Ir. Acad.}},
  volume={27},
  pages={69--138},
  year={1907},
}

@article{whidborne2007minimization,
  title={On the minimization of maximum transient energy growth},
  author={Whidborne, James F and McKernan, John},
  journal={{IEEE Trans. Autom. Control}},
  volume={52},
  number={9},
  pages={1762--1767},
  year={2007},
  publisher={IEEE}
}

@book{khalil2002nonlinear,
  title={{Nonlinear systems}},
  author={Khalil, H. K.},
  publisher={Upper Saddle River},
  year={2002}
}

@inproceedings{floquet1883equations,
  title={{Sur les {\'e}quations diff{\'e}rentielles lin{\'e}aires {\`a} coefficients p{\'e}riodiques}},
  author={Floquet, Gaston},
  booktitle={Ann. Sci. Ec. Norm. Super.},
  volume={12},
  pages={47--88},
  year={1883}
}

@article{butler1992three,
  title={Three-dimensional optimal perturbations in viscous shear flow},
  author={Butler, Kathryn M and Farrell, Brian F},
  journal={Phys. Fluids},
  volume={4},
  number={8},
  pages={1637--1650},
  year={1992},
  publisher={American Institute of Physics}
}

@article{davis1976stability,
  title={The stability of time-periodic flows},
  author={Davis, Stephen H},
  journal={Annu. Rev. Fluid Mech.},
  volume={8},
  number={1},
  pages={57--74},
  year={1976},
  publisher={Annual Reviews 4139 El Camino Way, PO Box 10139, Palo Alto, CA 94303-0139, USA}
}

@book{schimd2001stability,
  title={Stability and Transition in Shear Flows},
  author={Schimd, P J and Henningson, D S},
  year={2001},
  publisher={Springer}
}

@article{trefethen1993hydrodynamic,
  title={Hydrodynamic stability without eigenvalues},
  author={Trefethen, Lloyd N and Trefethen, Anne E and Reddy, Satish C and Driscoll, Tobin A},
  journal={Science},
  volume={261},
  number={5121},
  pages={578--584},
  year={1993},
  publisher={American Association for the Advancement of Science}
}

@article{xu2021non,
  title={Non-modal transient growth of disturbances in pulsatile and oscillatory pipe flows},
  author={Xu, Duo and Song, Baofang and Avila, Marc},
  journal={J. Fluid Mech.},
  volume={907},
  pages={R5},
  year={2021},
  publisher={Cambridge University Press}
}

@book{Boyd1994,
  author = {Boyd, S. and El Ghaoui, L. and Feron, E. and Balakrishnan, V.},
  title = {Linear Matrix Inequalities in System and Control Theory},
  publisher = {Society for Industrial and Applied Mathematics},
  year = {1994}
}

@article{Liu2020,
  author = {Liu, C. and Gayme, D. F.},
  title = {Input-output inspired method for permissible perturbation amplitude of transitional wall-bounded shear flows},
  journal = {Phys. Rev. E},
  volume = {102},
  pages = {063108},
  year = {2020}
}

@article{kalur2021nonlinear,
  title={Nonlinear stability analysis of transitional flows using quadratic constraints},
  author={Kalur, Aniketh and Seiler, Peter and Hemati, Maziar S},
  journal={Phys. Rev. Fluids},
  volume={6},
  number={4},
  pages={044401},
  year={2021},
  publisher={APS}
}

@ARTICLE{kalur2022estimating,
  author={Kalur, Aniketh and Mushtaq, Talha and Seiler, Peter and Hemati, Maziar S.},
  journal={IEEE Control Syst. Lett.}, 
  title={Estimating Regions of Attraction for Transitional Flows Using Quadratic Constraints}, 
  year={2022},
  volume={6},
  number={},
  pages={482-487},
  doi={10.1109/LCSYS.2021.3081382}}

@article{Linot_Schmid_Taira_2024, title={On the laminar solutions and stability of accelerating and decelerating channel flows}, volume={999}, DOI={10.1017/jfm.2024.709}, journal={J. Fluid Mech.}, author={Linot, Alec J. and Schmid, Peter J. and Taira, Kunihiko}, year={2024}, pages={A43}}

@article{schmid1994optimal,
  title={Optimal energy density growth in {H}agen--{P}oiseuille flow},
  author={Schmid, Peter J and Henningson, Dan S},
  journal={J. Fluid Mech.},
  volume={277},
  pages={197--225},
  year={1994},
  publisher={Cambridge University Press}
}

@INPROCEEDINGS{Monta2003,
  author={Montagner, V.F. and Peres, P.L.D.},
  booktitle={42nd IEEE Conference on Decision and Control}, 
  title={{A new {LMI} condition for the robust stability of linear time-varying systems}}, 
  year={2003},
  volume={6},
  number={},
  pages={6133-6138},
  doi={10.1109/CDC.2003.1272249}}

@ARTICLE{Jetto2009,
  author={Jetto, Leopoldo and Orsini, Valentina},
  journal={{IEEE Trans. Autom. Control}}, 
  title={{LMI} Conditions for the Stability of Linear Uncertain Polynomially Time-Varying Systems}, 
  year={2009},
  volume={54},
  number={7},
  pages={1705-1709},
  doi={10.1109/TAC.2009.2020645}}

@INPROCEEDINGS{chesi2004,
  author={Chesi, G. and Garulli, A. and Tesi, A. and Vicino, A.},
  booktitle={43rd IEEE Conference on Decision and Control}, 
  title={Parameter-dependent homogeneous {L}yapunov functions for robust stability of linear time-varying systems}, 
  year={2004},
  volume={4},
  number={},
  pages={4095-4100},
  doi={10.1109/CDC.2004.1429393}}

@article{gustavsson1986excitation,
  title={Excitation of direct resonances in plane {P}oiseuille flow},
  author={Gustavsson, L H{\aa}kan},
  journal={Stud. Appl. Math.},
  volume={75},
  number={3},
  pages={227--248},
  year={1986},
  publisher={Wiley Online Library}
}

@article{linot2025extractingdominantdynamicsunsteady,
      title={{Extracting dominant dynamics about unsteady base flows}}, 
      author={Alec J. Linot and Barbara Lopez-Doriga and Yonghong Zhong and Kunihiko Taira},
      year={2025},
      journal={Fluid Dyn. Res.},
    volume={57},
    pages={031401}
}

@article{wei2025nonlinear,
  title={Nonlinear input-output analysis of transitional shear flows using small-signal finite-gain $\mathcal{L}_p$ stability},
  author={Wei, Zhengyang and Liu, Chang},
  journal={Phys. Rev. Fluids},
  volume={10},
  pages={103903},
  year={2025}
}

@inproceedings{Lofberg2004,
  author = {Lofberg, J.},
  title = {{YALMIP}: A toolbox for modeling and optimization in {MATLAB}},
  booktitle = {Proceedings of the IEEE International Conference on Robotics and Automation},
  volume = {3},
  pages = {284--289},
  year = {2004}
}

@manual{Mosek2024,
  author = {{MOSEK ApS}},
  title = {The MOSEK optimization toolbox for MATLAB manual},
  year = {2024}
}

@article{kochnev2025stability,
  title={Stability analysis of thermohaline convection with a time-varying shear flow using the {L}yapunov method},
  author={Kochnev, Kalin and Liu, Chang},
  journal={arXiv preprint arXiv:2509.26545},
  year={2025}
}

\end{document}